\documentclass[12pt, letterpaper]{article}
\usepackage[top=1in, bottom=1.5in, left=1in, right=1in]{geometry}
\usepackage[utf8]{inputenc}
\usepackage{booktabs}
\usepackage{hyperref}
\usepackage{graphicx}
\usepackage{natbib}
\usepackage{colortbl}
\usepackage{amssymb}
\usepackage{aas_macros}

\usepackage{color}

\title{Enabling Deep All-Sky Searches of Outer Solar System Objects}
\author{Mario Juri\'c, R. Lynne Jones, J. Bryce Kalmbach, Peter Whidden, \\
Dino Bekte\v{s}evi\'c, Hayden Smotherman, Joachim Moeyens, Andrew J. Connolly, \\ Michele T. Bannister, Wesley Fraser, David Gerdes, Michael Mommert, \\ Darin Ragozzine, Megan E. Schwamb, and David Trilling}
\date{}

\begin{document}

\maketitle

\begin{abstract}
    A foundational goal of the Large Synoptic Survey Telescope (LSST) is to map the Solar System small body populations that provide key windows into understanding of its formation and evolution \citep{2008arXiv0805.2366I,2009arXiv0912.0201L}. This is especially true of the populations of the Outer Solar System -- objects at the orbit of Neptune $r > 30$~AU and beyond.

    In this whitepaper, we propose a minimal change to the LSST cadence that can greatly enhance LSST's ability to discover faint distant Solar System objects across the entire {\em wide-fast-deep} (WFD) survey area. Specifically, we propose that the WFD cadence be constrained so as to deliver least one sequence of $\gtrsim 10$ visits per year taken in a $\sim 10$ day period in any combination of $g, r$, and $i$ bands. Combined with advanced shift-and-stack algorithms \citep[e.g.][]{whidden+2019} this modification would enable a nearly complete census of the outer Solar System to $\sim 25.5$ magnitude, yielding $4-8$x more KBO discoveries than with single-epoch baseline, and enabling rapid identification and follow-up of unusual distant Solar System objects in $\gtrsim 5$x greater volume of space.

    These increases would enhance the science cases discussed in Schwamb et al. 2018 whitepaper, including probing Neptune's past migration history as well as discovering hypothesized planet(s) beyond the orbit of Neptune (or at least placing significant constraints on their existence). 
\end{abstract}

\section{White Paper Information}
Corresponding Author: Mario Juri\'c (mjuric@astro.washington.edu) 
\begin{enumerate} 
\item {\bf Science Category:} Taking an Inventory of the Solar System
\item {\bf Survey Type Category:} wide-fast-deep
\item {\bf Observing Strategy Category:}  an opportunity to significantly enhance a broad array of Solar System science programs by a limited modification of observing cadence.
\end{enumerate}

\section{Scientific Motivation}\label{sec:motivation}


The inventory of the Solar System is one of the four fundamental science themes of the Large Synoptic Survey Telescope (LSST). While LSST's impact will be significant across all populations, its transformative ability to discover and follow up faint objects at the outskirts of the Solar System sets it apart from any other existing or planned program. A survey strategy similar to the present baseline and including the Northern Ecliptic Spur (``NES'') would exhaustively map the outer Solar System, increasing the number of KBOs known to a limiting magnitude $r \approx 24.5$ by a factor of $10$ or more relative to today's numbers (Schwamb et al. 2018). Furthermore, a special ``deep drilling'' survey could sample the Kuiper Belt to $\approx 27$ (Trilling et al. 2018), but covering a significantly smaller area (few $10$~deg$^2$).

The implementation of this proposal would increase the number of feasibly discoverable Kuiper Belt Objects by factors of 4-8x over the entire WFD area. This would commensurately increase the signal-to-noise across all types of KBO population studies, opening the possibilities for identification of smaller sub-populations (especially the interesting resonant populations). It would also enhance the sensitivity to unusual objects by the same factor, making it more likely that objects such as the hypothesized planets beyond the orbit of Neptune are discovered (or their existence constrained).
\\


We stress that this proposal {\em should not be seen as an alternative} to Schwamb et al. NES proposal, but rather as an {\em enhancement}. The full coverage of the ecliptic remains {\em essential} for fully achieving the Solar System science goals. Nevertheless, the implementation of this proposal does not hinge on the adoption of Schwamb et al. 2018.

\subsection{High-level description}

We propose to add a constraint to the WFD cadence to ensure that at least once a year a sequence of visits (on order of $n=10$) is taken in either combination of $g$, $r$ and $i$ bands in a relatively short period of time (on order of $\Delta T=10$ days). Such a sequence would then be repeated once every year for the duration of the survey. If executed as described, for each pointing at the end of the survey there'd be ten sequences of 10 visits taken in a 10-day period, separated roughly by a year.

Having visits in such sequences would enable the application of state-of-the-art shift-and-stack algorithm to search for and detect moving objects in the distant Solar System (Kuiper Belt and beyond) throughout the WFD footprint, once for each sequence. Taking the sequences annually would ensure that the detected objects are re-detected at least once per year, enabling opposition-to-opposition linking and orbit estimation.



\subsection{Footprint -- pointings, regions and/or constraints}

This proposal has no strong constraints on the footprint; we would optimally prefer the sequence of pointings described here to be executed wherever the LSST observes.
\\

However, should that not be possible, we offer two fall-back scenarios:
\begin{itemize}
    \item Do not execute the proposed sequences in sky areas of high object density (i.e., the Galactic Plane). In these regions the stacking becomes challenging due to high stellar number density (and, depending on the algorithm, an increase in image differencing artefacts).
    \item Do not execute the sequences more than $\pm 30^{\circ}$ away from the ecliptic. Given the known KBO populations are largely concentrated in this band, this would preserve most of the science enabled by this proposal. However, we would lose the additional sensitivity to unusual objects and populations residing away from the ecliptic.
\end{itemize}

The details of the tradeoffs are further discussed in Section~\ref{sec:trades}.

\subsection{Image quality}

There are no special considerations regarding image quality.


\subsection{Individual image depth and/or sky brightness}

We assume the individual image depth will be close to existing LSST requirements. In general, deeper single-epoch images would be preferred, as they reduce the number of visits required to reach a given co-added depth. Fewer visits reduces the computational power needed for processing (further discussed in Sec~\ref{sec:trades}).

\subsection{Co-added image depth and/or total number of visits}

The number of KBOs that can be discovered relative to a strategy employing single-visits scales approximately linearly with the number of exposures taken in any given sequence (i.e., by taking a sequence of two visits we expect to find 2x more KBOs, with 5 visits 5x, etc.; note that this is also strongly dependent on the underlying KBO population distribution). To achieve a meaningfully signficant increase in the number of discovered KBOs, we propose a nominal factor of $n=10$ as the total number of visits per sequence.

We note that this proposal does not depend on intra-night revisits (as it does not rely on MOPS\footnote{\url{http://ls.st/LDM-156}} for linking and discovery).

\subsection{Distribution of visits over time}
\label{sec:distvisits}

When exploring the distribution of visits we attempt strike a balance between the depth of the discovery enabled by sequences of close-in-time visits, against the number of such sequences needed to enable re-discovery and orbit determination.

Here we propose to take one sequence of 10 visits, roughly equally distributed over a timespan of 10 days, once per year. All 10 visits do not have to be in the same band, but must be in the combination of $g$, $r$, and $i$ bands (where Solar System objects are easily detectable). This nominal distribution would increase the effective LSST depth for KBO discoveries by up to $\Delta m \eqsim 1.25$ mag, while providing sufficient number of detections (10) over a long enough temporal baseline (10 years) to determine the orbit.

The number of observations in each sequence affects the effective limiting magnitude, scaling approximately as $\Delta m \propto 1.25 \log(n)$, where $n$ is the number of visits in a sequence. The exact number of observations in a sequence is flexible; we have not identified any hard ``cliffs'' in the science as a function of depth. Because the objects need to be recovered each year to determine the orbits, it's important that the number of exposures in each sequence does not vary significantly (on order of $\sim 10\%$).


The length of each sequence affects the computational power needed to process the sequence, which scales as $\Delta T^2$ (Kalm`bach et al., submitted). It is therefore preferable to keep the sequences relatively short. We selected our baseline of 10 days based on our projections of algorithmic improvements and the computational power likely to be available for large-scale science analyses in LSST DR2 timeframe (as discussed in Sec~\ref{sec:computation}).

\subsection{Filter choice}
Solar System minor planets, visible from reflected sunlight, are brightest in the mid-optical wavelengths. We therefore prioritize observations in $g$, $r$ and $i$-bands. This filter choice maximizes the KBO science for the reasons discussed in Schwamb et al. (2018).

\subsection{Exposure constraints}
This proposal imposes no special minimum or maximum exposure time constraints.

\subsection{Other constraints}
None noted.               

\subsection{Estimated time requirement}


This proposal requires no additional time.

\subsection{Technical trades}
\label{sec:trades}

\subsubsection{What is the effect of a trade-off between your requested survey footprint (area) and requested co-added depth or number of visits}

The primary trade is between:
\begin{itemize}
    \item the average size (or distance) of discovered objects (dictated by the number of visits/coadded depth) and
    \item the understanding of the global distribution of objects throughout the orbital parameter space as well as maximizing the discovery space for unusual objects.
\end{itemize}

Should trades between these parameters be necessary, we prefer to reduce the area first, and depth second, in the following way:
\begin{itemize}
    \item Do not execute the sequences in sky areas of high object density (i.e., the Galactic Plane). In these regions the stacking becomes challenging due to high stellar number density (and, depending on the algorithm, an increase in image differencing artefacts).
    \item Deprioritize executing sequences away from the ecliptic. Given the known KBO populations are largely concentrated in a $\pm 30^{\circ}$ band around the ecliptic, this would preserve most of the discovery space enabled by this proposal. However, this carries a serious consequence of reducing the LSST's sensitivity to unusual objects and populations residing away from the ecliptic.
\end{itemize}

 \subsubsection{If not requesting a specific timing of visits, what is the effect of a trade-off between the uniformity of observations and the frequency of observations in time? e.g. a `rolling cadence' increases the frequency of visits during a short time period at the cost of fewer visits the rest of the time, making the overall sampling less uniform.}

This proposal explicitly asks to introduce a rolling-cadence type non-uniformity\footnote{We understand this type of cadence may also be beneficial for variability studies (e.g., variable stars and/or transients)}.

The length of each high-cadence sequence affects the computational power needed to process the sequence, which scales as $\Delta T^2$ \citep{whidden+2019}. It is therefore preferable to keep the sequences relatively short.

\subsubsection{What is the effect of a trade-off on the exposure time and number of visits (e.g., increasing the individual image depth but decreasing the overall number of visits)?}

This proposal would prefer longer exposure over more visits. This would allow for shorter sequence durations, quadratically reducing the computational power needed for the shift-and-stack algorithm (Section~\ref{sec:computation}).
 
 \subsubsection{What is the effect of a trade-off between uniformity in number of visits and co-added depth? }

The spacing of observations within a sequence is not strongly constrained, though it's preferable to be close to uniform (e.g., one visit per night, or two visits per night every other night, or three visits per night every other night, etc.). Extreme bimodality (such as five visits in one night, followed by a five visits 10 nights later) are disfavored (as it would be more difficult to associate the detections with certainty).

\subsection{Is there any benefit to real-time exposure time optimization to obtain nearly constant single-visit limiting depth?}

This proposal would not be adversely affected by such an optimization. It's not clear whether there would be any benefits.

\subsubsection{Are there any other potential trade-offs to consider when attempting to balance this proposal with others which may have similar but slightly different requests?\label{sec:snaps}}

None have been identified.

\section{Performance Evaluation}


We identify two metrics of interest relevant to this proposal: the total number of discovered distant Solar System objects, $N_{KBO, S}$, and the additional Solar System volume available to discovery.

\subsection{Number of Discoverable KBOs}

We begin by defining the ratio $f_{\rm S}$ of the number KBOs available for discovery given the proposed sequences, over what would be discovered given single-epoch detections:
\begin{equation}
    f_{\rm S} = \frac{N_{KBO, S}}{N_{\rm KBO, single-epoch}}
\end{equation}

Assuming the power-law distribution of the number of KBOs as a function of apparent magnitude of $\Sigma(m) \propto 10^{0.75 m}$ (following Petit et al. 2008), and the scaling of effective limiting magnitude as $\Delta m \propto 1.25 \log(n)$ where $n$ is the number of visits in a sequence, this factor can be shown to approximately scale as:
\begin{equation}
    f_{\rm S} = n^{0.9375}
\end{equation}
assuming the same depth for each exposure. That means that, given a sequence of 5 exposures in a given area of the sky, {\bf the LSST would become capable of finding $\approx 4.5$x more KBOs than with a single visit cadence}\footnote{This number may be smaller, up to a factor of two, as the KBO size distribution has been shown to exhibit a break right around the LSST single-epoch magnitude limit (e.g., Holman \& Fuentes 2009)}.

Then, the total number of discoverable objects, $N_{KBO, S}$, becomes equal to:
\begin{equation}
    N_{KBO, S} = \sum_{i \in {\rm pointings} } f_{\rm S,i}  N_{{\rm KBO, single-epoch, }i}
\end{equation}
where $N_{{\rm KBO, single-epoch}}$ is sourced from the LSST's Solar System model. This metric should be maximized across the survey.

\subsection{Discovery Volume}

The metric described in previous section can be made close to maximal by positioning most of the sequences in a band around the ecliptic. This, however, reduces the discovery space for {\em unknown and peculiar objects} found on odd orbits away from the ecliptic. We therefore introduce another metric to provide a counter-balance and retain that element of the scientific use case.

We define the {\em discovery volume}, $V_{H, S}$, for objects of fiducial absolute magnitude $H$ which scales as:
\begin{equation}
    V_{H, S} = 10^{\frac{3}{3} \Delta m} \, V_{H}
\end{equation}
relative to the single-epoch discovery volume $V_{H}$ given the additional $\Delta m$ of depth. In addition to the total number of discoveries, we also wish to maximize this discovery volume, summed up over all pointings in the survey.
\\

There is some room to trade between these two metrics, as well as to choose an design point for either of them. We identify no hard ``cliffs'' in either the number of objects or the discovery volume metrics under which improved science would be impossible (though, obviously, having both of these metrics close to their single-epoch values would be disappointing). For rough guidance, given our fiducial proposal of $n = 10$ observations in a sequence evaluated over the entire WFD footprint, the values of the two metrics are:
\begin{equation}
    N_{KBO,S} = 8.7 \, N_{KBO}
\end{equation}
and
\begin{equation}
    V_{H, S} = 5.6 \, V_{H}
\end{equation}. A reasonable fiducial goal may be to keep $N_{KBO, S} > 5$ and $V_{H, S} > 3$.

\section{Special Data Processing}
\label{sec:computation}

To discover additional distant Solar System Objects given the sequences proposed here requires the application of shift-and-stack techniques to search for moving sources below the detection limit of any individual image \citep{Gladman+1997, Allen+2001,Bernstein+2004,Heinze+2015}. These approaches are fundamentally different from the traditional techniques in that they assume a trajectory for an asteroid and align a set of individual images along that trajectory in order to look for evidence for a source. 

Recently, \cite{whidden+2019} have introduced a new shift-and-stack type computational technique for searching for faint moving sources in astronomical images. Starting from a maximum likelihood estimate for the probability of the detection of a source within a series of images, they developed a massively parallel algorithm for searching through candidate asteroid trajectories that utilizes Graphics Processing Units (GPU). This technique can search over $10^{10}$ possible asteroid trajectories in stacks of the order $10-15$ 4K x 4K images in under a minute using consumer grade GPUs, and has been successfully demonstrated using DECam data.

Using the timings from the \cite{whidden+2019} study and the fiducial numbers for the number of exposure and legth of the proposed sequence, we find that processing one year of LSST data would require approximately $3.5$M~GPU-hours. This equates to approximately 40-days of computing on present-day national petascale resources (the 2013-era Blue Waters supercomputer at NCSA, co-located with the LSST dataset). Assuming further hardware improvements (Moore's law is still valid for GPUs) and modest algorithmic improvements, this run time is likely to be an order of magnitude lower in the $\sim 2025$ timeframe. This will place it comfortably within the range of an computational cycle grant available to NSF researchers on NSF's computational resources.

\section{Acknowledgements}
 
 The authors thank the  Large Synoptic Survey Telescope (LSST) Project Science Team and the LSST Corporation for their support of LSST Solar System Science Collaboration's (SSSC) efforts. This work was supported in part by a LSST Corporation Enabling Science grant. The authors also thank the B612 Foundation, AURA, and the Simons Foundation for their support for workshops, hackathons, and sprints that lead to the development of this white paper. Elements of this work were enabled by the Solar System JupyterHub service at the University of Washington's DIRAC Institute (\url{http://dirac.astro.washington.edu}). This white paper has made use of NASA's Astrophysics Data System Bibliographic Services.

\bibliographystyle{aasjournal}
\bibliography{references} 
\end{document}